\documentclass[aps,prl,twocolumn]{revtex4-2}
\usepackage{amsmath}
\usepackage{amssymb}
\usepackage{graphicx}
\usepackage{xcolor}
\usepackage[hidelinks]{hyperref}
\usepackage{url}
\usepackage{fancyhdr}

\fancypagestyle{firstpage}{ 
	
	\fancyhead{\fontsize{8}{13}\selectfont PHYSICAL REVIEW LETTERS {\bf 132}, 057201 (2024)}
	\fancyhead[L]{}
	\fancyhead[R]{}
	\fancyfoot[C]{\thepage}
}
\fancypagestyle{plain}{ 

\fancyfoot[C]{\thepage}}
\newcommand{\vp}{\varphi}
\newcommand{\cc}{\bar}
\newcommand{\ii}{\mathrm{i}}
\newcommand{\R}{\mathbb{R}}
\newcommand{\C}{\mathbb{C}}

\def\footnoterotatedperspective{
There is some freedom in the way transformation~\eqref{eq:transformation} relates original state variables $x_j$ with the dynamical variables $Q,y,s$. 
One can always consider variables rotated in the complex plane: $\Phi = e^{-\ii\alpha}Q$, $\lambda = e^{-\ii\beta}y$, 
in which case the transformation is expressed as: $x_j = e^{\ii\alpha}\Phi + e^{\ii\beta}\lambda \frac{\xi_j}{1+s\xi_j}$, 
and the dynamical equations as: $\dot{\Phi} = e^{\ii\alpha} a \Phi^2 + b \Phi + e^{-\ii\alpha}c$, $\dot{\lambda} = (b+e^{\ii\alpha}2a\Phi)\lambda$, $\dot{s} = -e^{\ii\beta}a\lambda$, 
cf. Eqs. ~(31) in~\cite{pietras_cestnik_pikovsky_2023} where $\alpha = \frac{\pi}{2}$ and $\beta = -\frac{\pi}{2}$. 
}
\def\footnotereseting{
When $Q$ is diverging $Q\to\infty$, one can reset the variables according to the initial condition~\eqref{eq:ic} again: $Q \mapsto 0,\ y\mapsto 1,\ s \mapsto 0$ and $\xi_j \mapsto x_j$. 
}

\begin{document}

\title{Integrability of a globally coupled complex Riccati array: \\  quadratic integrate-and-fire neurons, phase oscillators and all in between}

\author{Rok Cestnik}
\email{rok.cestnik@math.lth.se}
\affiliation{Centre for Mathematical Science, Lund University, Sölvegatan 18, 22100, Lund, Sweden}

\author{Erik A. Martens}
\affiliation{Centre for Mathematical Science, Lund University, Sölvegatan 18, 22100, Lund, Sweden}

\date{\today}

\begin{abstract}
We present an exact dimensionality reduction for dynamics of an arbitrary array of globally coupled complex-valued Riccati equations. It generalizes the Watanabe-Strogatz theory [Phys. Rev. Lett. 70, 2391 (1993)] for sinusoidally coupled phase oscillators and seamlessly includes quadratic integrate-and-fire neurons as the real-valued special case. This simple formulation reshapes our understanding of a broad class of coupled systems -- including a particular class of phase-amplitude oscillators -- which newly fall under the category of integrable systems. Precise and rigorous analysis of complex Riccati arrays is now within reach, paving a way to a deeper understanding of emergent behavior of collective dynamics in coupled systems. 
\end{abstract}

\maketitle
\thispagestyle{firstpage}

The study of complex systems often involves describing them using a few relevant variables known as order parameters. Such dimensionality reduction techniques are highly useful but challenging to discover, and they may not exist for every system.
In this regard, Watanabe and Strogatz~\cite{watanabe_strogatz_1993} (WS) made significant progress by demonstrating how globally coupled phase oscillators can be effectively described using only three order parameters. This reduction arises from a specific subclass of complex Riccati equations~\cite{watanabe_strogatz_1994,Pikovsky2008} that govern the dynamics of such oscillatory arrays. The macroscopic variables are parameters of a Möbius transform that relates the dynamical state variables to constants determined by the initial states of the oscillator array~\cite{marvel_mirollo_strogatz_2009}. Remarkably, Ott and Antonsen~\cite{ott_antonsen_2008} showed that for large ensembles of nonidentical oscillators, the dynamics even collapses to an effectively 2D manifold. 
These descriptions have proven immensely useful in studying the collective dynamics of coupled oscillatory systems~\cite{Pikovsky2015} and found many applications, particularly in neuroscience~\cite{Bick2020,Coombes2018next}. 

In this Letter, we present a framework that generalizes these dimensionality reductions to a larger class of systems: an arbitrary array $j=1,...,N$ of globally forced complex Riccati equations, 


\begin{equation}
	\dot{x}_j = a x_j^2 + b x_j + c\;,\qquad a,b,c\in\C \,,
\label{eq:riccati}
\end{equation}
where all $a,b,c \in \C$ can be arbitrary complex functions of time and $x_j \in \C$ can start with arbitrary complex values. 
The choice for $a,b,c$ allows selecting from a range of coupled oscillator systems. Certain choices reproduce known models, such as the quadratic integrate-and-fire neurons (QIF) and phase oscillators; but the 
possibilities go far beyond that and cover a broad class of 2D systems. 


The flow of array variables $x_j$ in~\eqref{eq:riccati} is given by the following Möbius transformation: 
\begin{equation}
	x_j = Q + \frac{y\, \xi_j}{1+s\, \xi_j}\,,
\label{eq:transformation}
\end{equation}
where the three global complex-valued variables $Q,y,s$ evolve according to:
\begin{subequations}
\begin{align}
	\dot{Q} &= a Q^2 + b Q + c \label{eq:Q}\,,\\
	\dot{y} &= (b+2aQ)y \label{eq:y}\,,\\
	\dot{s} &= -ay \label{eq:s}\,,
\end{align}
\label{eq:Qys}
\end{subequations}
and $\xi_j \in \C$ are constants determined by the initial values of the oscillator array variables~\footnote{\footnoterotatedperspective} (cf. Eqs.~(4) in~\cite{Goebel1995} and Eqs.~(15-17) in~\cite{Lohe2019}).
Since these equations completely generate the flow of the original system~\eqref{eq:riccati}, this implies that the dynamics of Eqs.~\eqref{eq:riccati} is effectively six dimensional. See Supplemental Material~\cite{SM} for a proof of this result.

{\it Choosing initial conditions. }
Since we started with an array of $N$ variables $\{x_j\}$ and now we describe the system with the three macroscopic variables $Q,y,s$ and $N$ constants $\{\xi_j\}$, we have some freedom in choosing the relationship between these variables. An additional constraint of choice is set on the initial values, that defines the exact relationship between $x_j$ and $Q,y,s,\xi_j$. 
We present two convenient options here. 
{\it (I) Identity conversion:}
a straightforward way to determine initial conditions is requiring that variables $x_j$ initially coincide with constants $\xi_j$ (cf. Eq.~(3.7) in~\cite{watanabe_strogatz_1994}) 
\begin{equation}
	Q(0) = 0\,,\quad y(0) = 1\,,\quad s(0) = 0\,,\quad \xi_j = x_j(0)\,.
\label{eq:ic}
\end{equation}
{\it (II) Möbius conversion:}
in general, the relation between $x_j$ and $\xi_j$ is a Möbius transform. We here use 
\begin{equation}
	Q(0) = \ii\,, \ y(0) = -2\ii\,, \ s(0) = 1\,, \ \xi_j = \frac{\ii-x_j(0)}{\ii+x_j(0)}\,.
\label{eq:ic_alt}
\end{equation}
In different situations different initial condition choices are more appropriate - we will see examples of both options in the later examples. 
Other options for initial constraints are possible as well, e.g., $\sum_j \xi_j = 0$ (cf. Eq.~(4.12) in~\cite{watanabe_strogatz_1994}), see Supplemental Material~\cite{SM} for more details. 


{\it Special case of real-valued arrays. }
We can consider the special case of real coefficients and real initial conditions: $a,b,c,x_j(0) \in \mathbb{R}$. 
Consequently, the flow of variables is real-valued, $x_j(t)\in\R$ for all $t\geq 0$. 
The dynamics is then three dimensional, which is easiest to see with constraint~\eqref{eq:ic} which make $Q,y,s$ initially real, and since they follow Eqs.~\eqref{eq:Qys} with real coefficients, they remain real with evolution. 
An example of this special case is a globally coupled array of identical QIF neurons~\cite{QIF_1986,Izhikevich_2007}. Individual voltages $x_j(t)$ obey
\begin{equation}
	\dot{x}_j = x_j^2+I\;, \qquad \text{if }\ x_j > x_\text{thr} \ \text{ then }\ x_j \mapsto x_\text{reset}\,,
\label{eq:QIF}
\end{equation}
where the voltage threshold and reset values are: $x_\text{thr} = \infty$ and $x_\text{reset} = -\infty$. 
The current $I$ can have a constant component $I_0$ associated with intrinsic neuronal dynamics, but it can also represent external forcing (even noisy) or global coupling, e.g., the input current generated by $N$ globally coupled QIF neurons with pulses $P(u)$ is expressed as: 
\begin{equation}
	I = I_0 + \frac{\epsilon}{N} \ \sum_{j=1}^N P\left(1/x_j\right)\,.
\label{eq:QIF_I}
\end{equation}
Within our formalism, the voltage spikes naturally occur when the denominator $1+s\xi_j$ in \eqref{eq:transformation} crosses zero. As a result, the voltage in that instance reaches $+\infty$ upon which it is reset to $-\infty$. What has to be considered, however, is that depending on the chosen initial constraint, $Q$ might also diverge. Indeed, this is the case if one chooses initial conditions according to~\eqref{eq:ic}: $Q(0) = s(0) = y(0)-1 = 0$ in which case additional resetting of variables would be needed~\footnote{\footnotereseting}. 
However, diverging variables and additional resetting can be avoided by simply choosing the appropriate initial conditions~\eqref{eq:ic_alt}: $Q(0) = \ii,\ y(0) = -2\ii,\ s(0) = 1$. The constants $\xi_j$ then relate to $x_j$ via a Möbius transform: $\xi_j = \frac{\ii-x_j(0)}{\ii+x_j(0)}$. Since $x_j(0)$ take real values, the constants $\xi_j$ are unitary $|\xi_j| = 1$, and can thus be described by their argument $\psi_j$: $\xi_j = e^{\ii\psi_j}$.  

\begin{figure}[h!]
\includegraphics[width=1.0\columnwidth]{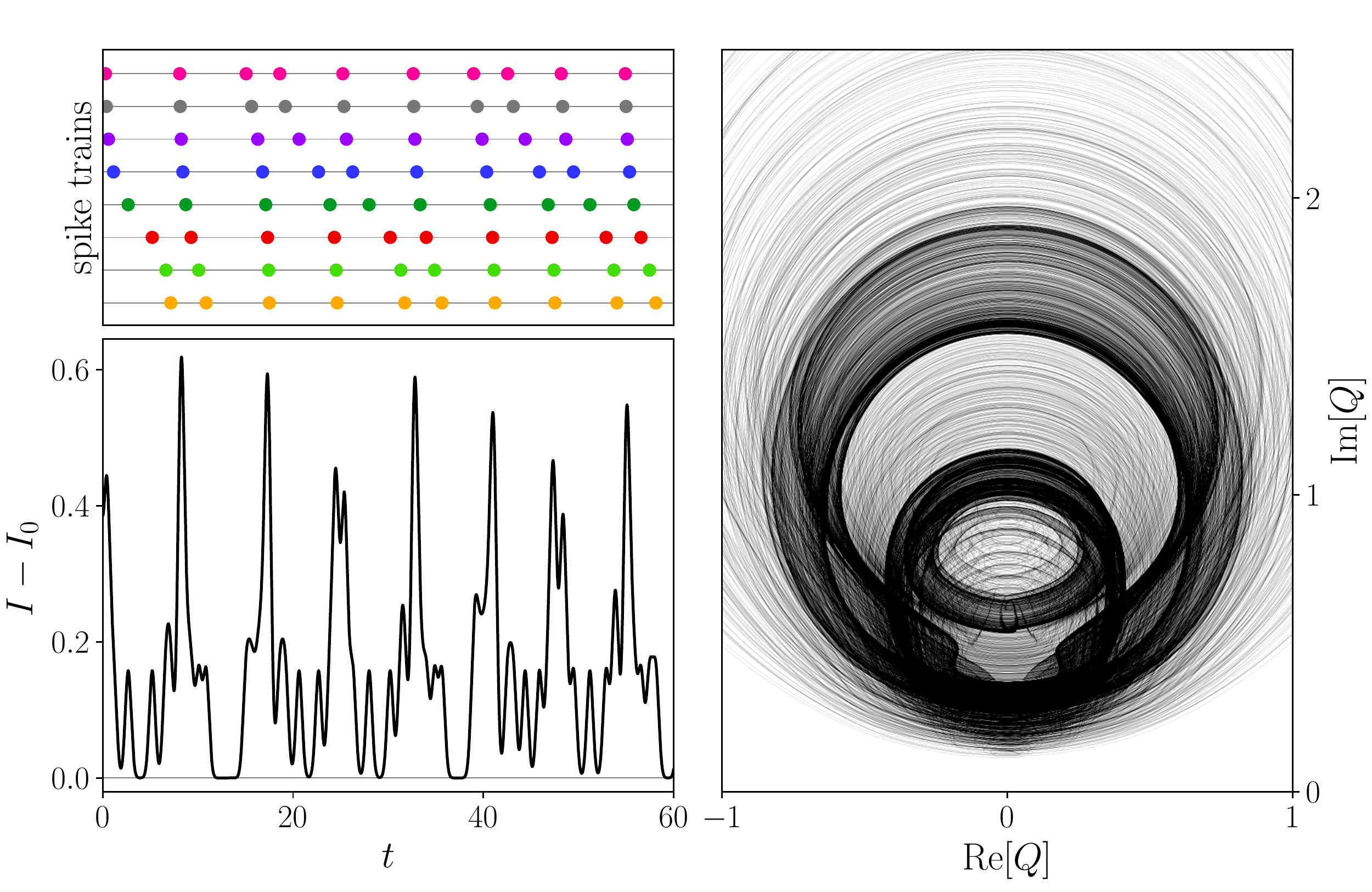}
	\caption{$N=8$ quadratic integrate-and-fire neurons with global coupling interacting via Gaussian pulses. The dynamics is exactly described by the low dimensional Eqs.~\eqref{eq:QIF_reduction}. Left panel: time series of the input current $I(t)$, alongside the individual neurons' firing events (top). Right panel: trajectory of the macroscopic variable $Q(t)$ determining voltages via \eqref{eq:transformation_QIF}. }
\label{fig:real_qif}
\end{figure}

For this specific case of $a,b,c,x_j \in \R$ and initial constraint~\eqref{eq:ic_alt} the following simplification is true:
\begin{equation}
	y = -(Q-\cc Q)s\,,
\end{equation}
which reduces the dynamical equations~\eqref{eq:Qys} to
\begin{subequations}
\begin{align}
	\dot{Q} &= aQ^2+bQ+c\,,\\
	\dot{\zeta} &= -\ii a(Q-\cc Q) = 2a\text{Im}[Q]\,,
\end{align}
\label{eq:QIF_reduction}
\end{subequations}
where $\zeta \in \mathbb{R}$ is the argument of $s = e^{\ii\zeta}$. Note that these dynamics are three dimensional since $Q$ now takes on complex values (even though $x_j \in \mathbb{R}$). 
The transformation~\eqref{eq:transformation} is reduced to:
\begin{equation}
	x_j \ = \ Q-(Q-\cc Q) \frac{e^{\ii(\psi_j+\zeta)}}{1+e^{\ii(\psi_j+\zeta)}} \ = \ \cc Q + \frac{(Q-\cc Q)}{1+e^{\ii(\psi_j+\zeta)}}\,.
\label{eq:transformation_QIF}
\end{equation}
The variable $Q$ remains bounded for all times, no additional resetting is needed and  the spikes occur when the denominator $1+e^{\ii(\psi_j+\zeta)}$ crosses 0, i.e., when $\zeta = \pi-\psi_j$. 
A numerical example of this special case is shown later in Fig.~\ref{fig:real_qif}. 
One can arrive at the same dynamics by transforming the QIF neurons into theta neurons via the transformation $x_j = \tan(\theta_j/2)$ and then employing the Watanabe-Strogatz theory for phase oscillators~\cite{watanabe_strogatz_1993,watanabe_strogatz_1994,marvel_mirollo_strogatz_2009}. For more on how our formalism relates to the Watanabe-Strogatz theory for phase oscillators see Supplemental Material~\cite{SM}, which includes Refs.~\cite{Pikovsky2011,strogatz2018nonlinear}. 

Such a description of QIF neurons has already been considered in the continuum limit $N\to\infty$, cf. Eqs.~(42) in~\cite{pietras_cestnik_pikovsky_2023}. If additionally the distribution of $x_j$ variables is Cauchy-Lorenztian, the dynamics further simplifies and variable $Q$ completely determines the macroscopic state, its real and imaginary component correlating to the firing rate and mean voltage~\cite{Montbrio2015}.


{\it Examples. }
We now present some specific systems where the dimensionality reduction can be applied and with numerical simulations validate its exactness. First (I), a known and relatable example of pulse-coupled real-valued QIF neurons. It is known that this system possesses low-dimensional dynamics since the QIF model can be transformed into a theta neuron to which the WS theory applies. Our formalism provides not only a new perspective of this fact, but also justifies the voltage resetting at infinity by viewing the model as a limiting case of the extended complex model. 
Next, we show two examples which generalize known models to complex values. 
Example (II), the complex generalization of the QIF model, where we simply allow the ``voltage'' variables to attain complex values. And example (III) concerns a complex-valued generalization of phase oscillators, specifically we choose overdamped Josephson junctions. 
An additional example is found in Supplemental Material~\cite{SM} of how our description applies to infinite ensembles in the thermodynamic limit, and how particular integrals can simplify with set initial conditions.  

{\it Example (I): real QIF model.}
We consider $N=8$ excitable QIF neurons~\eqref{eq:QIF},\eqref{eq:QIF_I} with $I_0 = -0.001$, interacting via Gaussian pulses $P(u) = \sqrt{\sigma/\pi} \exp(-\sigma u^2)$ where $\sigma = 5$ and coupling strength $\epsilon = 2.3$. Initial conditions are $\{x_j(0)\} = \{-(N-1)/2+j\}$, $j = 1,...,N$. These parameters yield chaotic dynamics, see Fig.~\ref{fig:real_qif}. We integrate the reduced three dimensional system using Eqs.~\eqref{eq:QIF_reduction} and compare the resulting trajectories with the ones obtained from $N=8$ coupled voltage equations~\eqref{eq:QIF}, we see an exact overlap, as expected. 

For ensembles of pure phase oscillators the Kuramoto order parameter is defined as $Z_1 = 1/N\sum_{j=1}^N e^{\ii\vp_j}$ where its amplitude $|Z_1|$ quantifies the order in the system. This is easily generalized to the full complex plane by simply invoking the mean~\footnote{Higher order parameters generalize to higher moments: $Z_n = \frac{1}{N}\sum_{j=1}^N x_j^n$.}:
\begin{equation}
	Z_1 = \frac{1}{N}\sum_{j=1}^N x_j\,,
\label{eq:order_parameter}
\end{equation}
which is neatly expressed with dynamical variables $Q,y,s$ and constants $\xi_j$,
\begin{equation}
	Z_1 = Q + y\frac{1}{N}\sum_{j=1}^N \frac{\xi_j}{1+s\xi_j}\,.
\label{eq:order_parameter_sum}
\end{equation}
The sum $1/N \sum_j \xi_j/(1+s\xi_j)$ can be seen as a function constant in time, that holds the information of initial conditions and only depends on the variable $s$. 

\begin{figure}[h!]
\includegraphics[width=1.0\columnwidth]{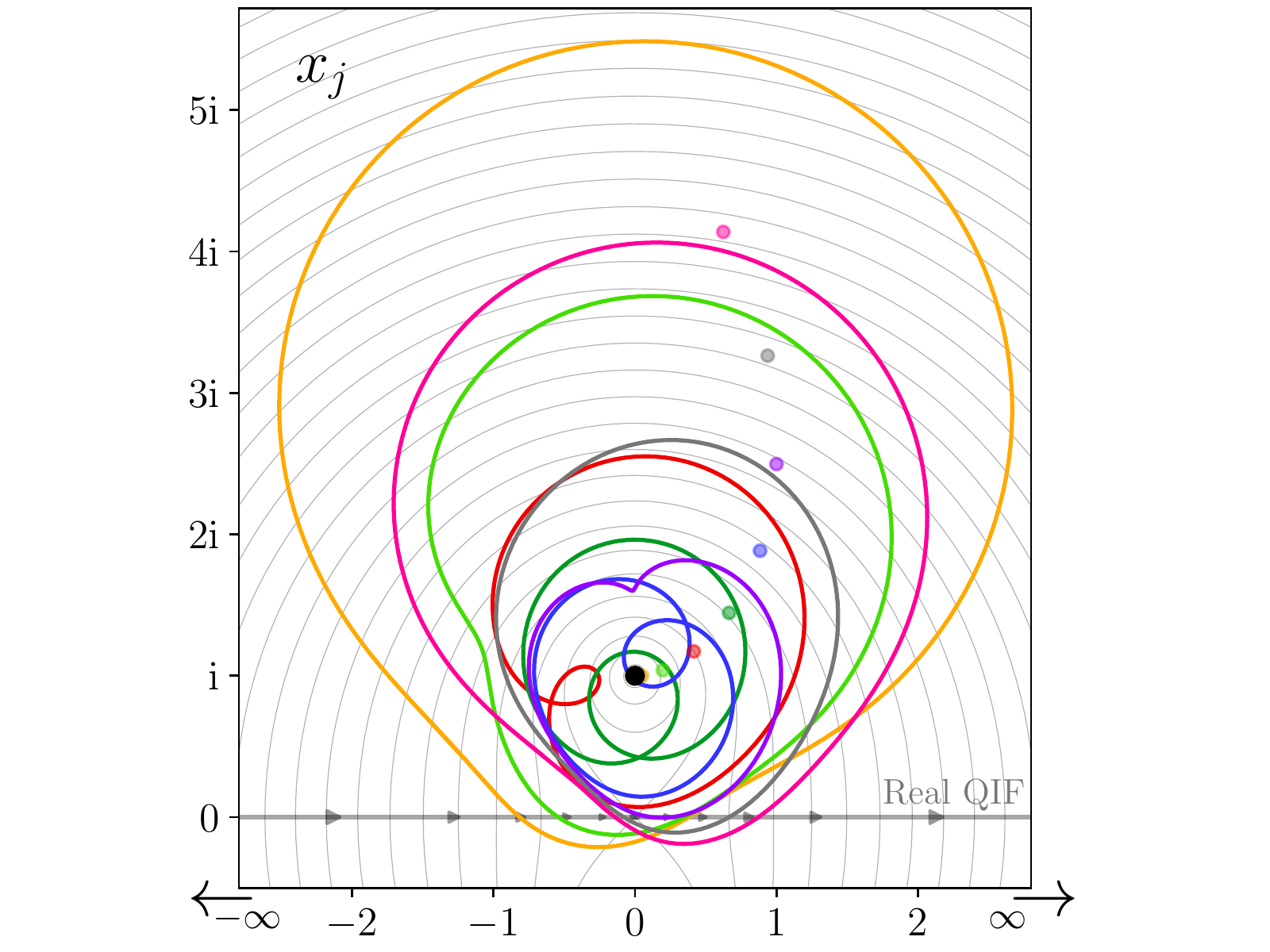}
	\caption{Complex generalization of coupled quadratic integrate-and-fire neurons~\eqref{eq:qif}. Unlike the real QIF model, the voltage resetting in this complex generalization is redundant since the dynamics everywhere (except the special case of real line with real coupling) loops back and stays finite; see also Example (I) in Fig.~\ref{fig:real_qif} where voltages diverge, but the resetting naturally results from the transformation~\eqref{eq:transformation}. $N=8$ units settle into periodic motion with a nontrivial limit cycle. Trajectories of oscillators on the limit cycle in the complex plane are depicted with colored lines. Initial conditions are marked with (color-coded) points. The fixed point $x_0^{+}$ is emphasized with a black dot. }
\label{fig:complex_QIF}
\end{figure}

{\it Example (II): complex QIF model.}
Let us consider a simple generalization of the QIF neurons to the complex plane. If the voltages $x_j$ start off the real axis, then they never diverge and there is no need for resetting conditions in~\eqref{eq:QIF}. Let us consider such generalized QIF neurons, globally coupled via the first moment $Z_1$~\eqref{eq:order_parameter},
\begin{equation}
	\dot{x}_j = x_j^2+I_0 \ \ + \epsilon \left(Z_1-x_0^{+}\right )\,,
\label{eq:qif}
\end{equation}
where $I_0$ is the intrinsic input current and $x_0^{+}$ the positive-imaginary fixed point of individual neurons (which can be absorbed in the current $I_0 \mapsto I_0-\epsilon x_0^{+}$). 
In the form of the initial Riccati equation~\eqref{eq:riccati} this model translates to the parameters: $a=1$, $b=0$, $c=I_0+\epsilon(Z_1-x_0^{+})$. 
On the real line the behavior of an individual unit $x_j$ tends towards infinity in finite time and hence one needs a reset rule~\eqref{eq:QIF} as well as an  implementation of a pulse during that event. However, if the dynamics occurs instead in the complex plane, the trajectory of this unit naturally oscillates around a fixed point. 
The uncoupled system is time-reversible, invariant under the involution: $\text{Re}[x_j] \mapsto -\text{Re}[x_j]$, $t\mapsto-t$, which allows for a continuum of closed orbits symmetric with respect to the imaginary axis if $I_0 > 0$, see Supplemental Material for details~\cite{SM}. 
Indeed, we find two center fixed points of the single unit dynamics: $x_0^{\pm} = \pm \ii \sqrt{I_0}$.
When we couple several units together, they remain oscillatory. In our example we use $N=8$ oscillators with $I_0 = 1$, $\epsilon = -5$ and initial conditions: $\{x_j(0)\} = \{x_0^{+}+\frac{j^2}{20} \exp(\ii\frac{\pi}{2N}(j-1))\}$, $j=1,...,N$. The dynamics settles into periodic motion with a nontrivial limit cycle, as shown in Fig.~\ref{fig:complex_QIF}. 

\begin{figure}[h!]
\includegraphics[width=1.0\columnwidth]{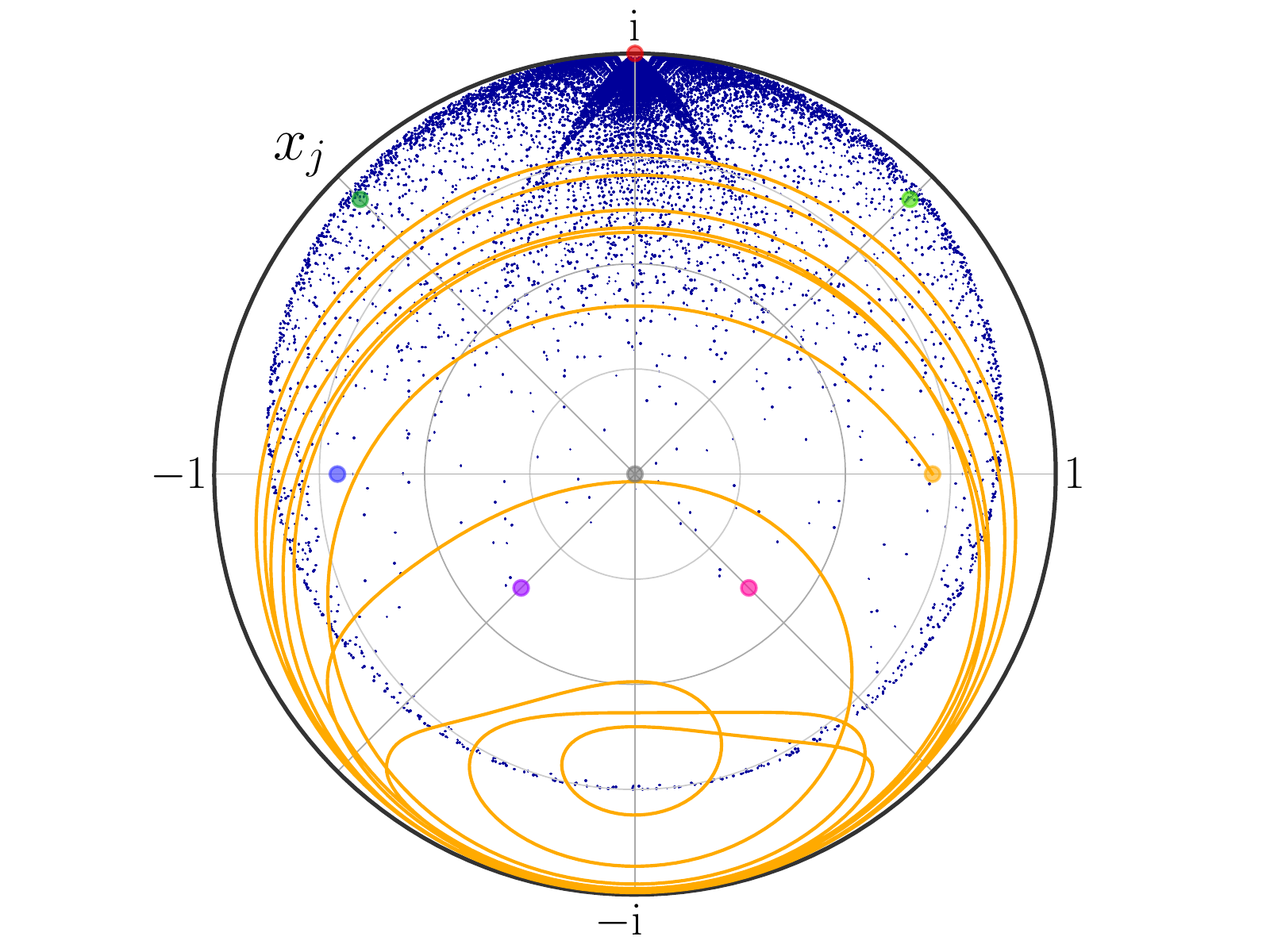}
	\caption{Array of $N=8$ Josephson junctions \eqref{eq:jj} with complex initial conditions inside the unit disk (colored points) generalize pure phase oscillators bound on the unit circle, see Fig.~4(c) in~\cite{watanabe_strogatz_1994} and Fig.~2 in~\cite{cestnik_pikovsky_2022}. Trajectory of one oscillator in the complex plane is shown in orange. Small blue scatter points depict the value of $Z_1$ on the Poincar\'e section where the oscillator on the unit circle passes phase $\pi/2$. }
\label{fig:jj_finite}
\end{figure}

{\it Example (III): complex generalization of phase oscillators.} 
Now let us generalize phase oscillators to include a free amplitude. 
Consider the phase dynamics equation in complex exponential form:
\begin{equation}
	\dot{x}_j = -\cc h x_j^2 + \ii \omega x_j + h\,,
\end{equation}
but allow that oscillators have amplitude $r_j$ different from 1: $x_j = r_j e^{i\vp_j}$. This choice results in a special family of phase-amplitude oscillators:
\begin{subequations}
\begin{align}
	\dot{\vp}_j &= \omega-\left(\frac{1}{r_j}+r_j\right)\text{Im}[he^{-i\vp_j}]\,,\\
	\dot{r}_j &= \left(1-r_j^2\right)\text{Re}[he^{-i\vp_j}]\,.
\end{align}
\label{eq:fi_r}
\end{subequations}
Note that $r_j = 1$ defines an invariant subspace that oscillators cannot cross: oscillators that start on the inside of the unit disk stay inside the disk forever. 
For this example we consider a complex generalization of coupled Josephson junctions (cf. Eq.~(3.16) in~\cite{watanabe_strogatz_1994}),
\begin{equation}
	a = -c = 0.75\;,\quad b = \ii-0.7\ii\ \text{Im}[Z_1]\,,
\label{eq:jj}
\end{equation}
where $Z_1$ is the generalized Kuramoto order parameter~\eqref{eq:order_parameter}.
We use $N = 8$ units with initial conditions: $\{x_j(0)\} = \{-\ii \sin(\frac{\pi}{N} j) \exp(\ii \frac{2\pi}{N} j)\}$, $j = 1,...,N$. Just like in the case of pure phase oscillators, for this parameter choice we observe chaos, see Fig.~\ref{fig:jj_finite}, cf. Fig.~4(c) in~\cite{watanabe_strogatz_1994} and Fig.~2 in~\cite{cestnik_pikovsky_2022}.


{\it Discussion.}
Our study presents a novel low-dimensional description that generalizes the well-known WS~\cite{watanabe_strogatz_1993,watanabe_strogatz_1994} theory to arbitrary arrays of complex Riccati equations, and so includes the real QIF model as a limiting case. This exact formalism enables the consideration of a whole new class of complex oscillatory models, including a special case of phase-amplitude oscillators. To showcase its correctness and applicability, we provide numerical simulations for several interesting examples.

The new formalism is effectively six dimensional, as compared to the three dimensional WS theory -- this is not surprising as the generalization from real $\mathbb{R}$ to complex $\mathbb{C}$ variables implies twice the dimensionality. However, it should be noted that the generalization does not simply involve allowing the dynamical variables to take complex values; rather, the equations we obtain are fundamentally different from those described by WS theory. What is common with the WS theory is that the overarching motif is the Möbius transform between initial values of the dynamical variables $x_j$ and the constants $\xi_j$. As was explored later~\cite{marvel_mirollo_strogatz_2009}, cross-ratios of dynamical variables $C_j = \frac{(x_j-x_{j+2})}{(x_j-x_{j+3})}\frac{(x_{j+1}-x_{j+3})}{(x_{j+1}-x_{j+2})}$ are invariant under the Möbius transform and thus constants of motion. Where in the WS context these cross-ratios are real (even though $x_j = e^{\ii\vp_j} \in \mathbb{C}$), here they can be complex: $C_j \in \mathbb{C}$. Just as was shown in~\cite{marvel_mirollo_strogatz_2009}, there are $N-3$ independent ratios $C_j$, $j = 1,...,N-3$. Since the initial problem contains $N$ complex variables~\eqref{eq:riccati}, and there are $N-3$ complex constants of motion, this leaves three complex variables to describe dynamics~\eqref{eq:Qys}, thus confirming that the description is six dimensional.  

For real-phase oscillators in the thermodynamic limit the inclusion of heterogeneity~\cite{ott_antonsen_2008,skardal_2018} or noise~\cite{toenjes_pikovsky_2020} leads to terms of complex-valued effective frequencies. Our description~\eqref{eq:Qys} is clearly applicable to describe those scenarios as well, as we have shown in previous work~\cite{cestnik_pikovsky_2022b, pietras_cestnik_pikovsky_2023}. This remarkable equivalence between noise and heterogeneity, and complex valued frequencies can be studied further with our approach. 

The generalization to complex numbers is substantial and provides room for qualitatively different dynamics. A complex extension of the Kuramoto model has recently been explored in~\cite{complexified_Kuramoto}; but our description applies to a much broader class of coupled systems described by complex Riccati arrays defined in Eqs.~\eqref{eq:riccati}. This unlocks a whole spectrum of 2D dynamical systems, including a special case of phase-amplitude oscillators~\eqref{eq:fi_r} we showcased here. The examples considered here are simply complex generalizations of known models: Example (II) a generalization of QIF and Example (III) a generalization of phase oscillators. One can explore systems that go beyond just generalizing known models to complex initial conditions, and really consider complex units with complex coupling, thus tapping into the rich 2D dynamics of intrinsic units defined by Riccati equations~\eqref{eq:riccati}.  

Moreover, the new description provides a fresh perspective and insights on the relationship between phase oscillators and QIF neurons. In fact, the voltage resetting in QIF neurons arises naturally in our framework, providing additional motivation for its use in studying neuronal dynamics and development of new models.

Our findings have significant implications for both theoretical and practical research. The new description opens up many avenues for investigating the dynamics of complex systems. 
Several ideas for future work directly follow from our framework, and more are expected to arise from the research community. We briefly outline three ideas here. 
(A) It is well known that the general Riccati equation can be transformed into a linear second order ODE~\cite{ince1956ordinary}, which means that our formalism applies there as well. A more detailed study will be performed in a forthcoming work. 
(B) Most likely similar descriptions exist for higher dimensional systems as well; both for systems with larger spatial dimensionality~\cite{chandra2019continuous}, $x_j \in\ \mathbb{R}^n$, $n>1$, such as the generalizations to vectors~\cite{lohe_2018} and matrices~\cite{Lohe2019}, as well as allowing for states $x_j$ that belong to higher number systems, such as quaternions or octonions. 
(C) For the special case of phase oscillators and QIF neurons, in the thermodynamic limit $N\to\infty$ a particular initial state leads to an additional dynamical reduction as described by Ott and Antonsen~\cite{ott_antonsen_2008}. One would expect that similar reductions can be found for the more general case of $a,b,c,x_j\in\C$. It would be interesting to know which states allow for additional reductions and what are the reduced dynamics. 
(D) Throughout this work we strictly considered identical oscillators, i.e., at all times every oscillator felt the same global forcings exerted by $a,b,c$. Thus, the oscillators only differed by their states $x_j$ which are determined by the initial conditions. 
However, one may consider adding heterogeneity in the forces by assuming that $a,b,c$ in some way differ between oscillators. Just like Ott and Antonsen incorporated Cauchy-Lorentzian inhomogeneities in frequencies~\cite{ott_antonsen_2008}, one can add them to our formalism in the thermodynamic limit as well. It is even likely that particular heterogeneity can be incorporated into finite arrays.

{\it Acknowledgments. }
We thank Arkady Pikovsky and Steven Strogatz for useful discussion and comments. We gratefully acknowledge financial support from the Royal Swedish Physiographic Society of Lund and the DFG (Grant No. PI 220/21-1).

\end{document}